\documentclass[twocolumn,prb]{revtex4}
\usepackage{graphicx}
\usepackage{dcolumn}
\usepackage{bm}
\usepackage{amsmath}

\begin{document}

\preprint{}
\title{Spin diffusion and magnetoresistance in ferromagnet/topological-insulator junctions }
\author{Takehito Yokoyama$^{1}$ and Yaroslav Tserkovnyak$^{2}$}
\affiliation{$^1$Department of Physics, Tokyo Institute of Technology, Tokyo 152-8551, Japan\\
$^2$ Department of Physics and Astronomy, University of California, Los Angeles, California 90095, USA}
\date{\today}

\begin{abstract}
We study spin and charge diffusion in metallic-ferromagnet/topological-insulator junctions. 
The diffusive theory is constructed for the coupled transport of the spin-dependent electron densities in the ferromagnet and the charge density in the topological insulator.
The diffusion equations for the coupled transport are derived perturbatively with respect to the strength of the interlayer tunneling. We analytically calculate spin accumulation in the ferromagnet and junction magnetoresistance associated with a current bias along the interface. It is found that due to the helical spin texture of the surface Dirac fermion, the spin accumulation  and the junction magnetoresistance depend on the angle between the magnetization and the current induced spin polarization on the surface of the topological insulator.

\end{abstract}

\pacs{PACS numbers:75.75.+a, 73.20.-r, 75.50.Xx, 75.70.Cn}
\maketitle




%

%



\section{Introduction}
\label{introduction}

Recent discovery of three-dimensional (3D) topological insulators (TI's) offers a new state of matter that is topologically distinct from the conventional band insulators.~\cite{Moore,Fu0,Schnyder,Qi2,Hasan,Qi1} Surface states of the strong TI are topologically protected and immune to small perturbations that respect time-reversal symmetry. At low energies, these states are described by Dirac fermions and hence exhibit strong spin-orbit interactions and intriguing Berry-phase phenomena. 
 
Since the electron's spin and momentum on the surface of a TI are essentially interlocked, TI's offer a promising arena for developing spintronics. In particular, there have been an intense interest in coupling TI's with ferromagnets (F's). In previous works, the focus has been on the surface properties of the TI's that are standalone or exchange coupled to the insulating F's,\cite{Culcer,Pesin,Yokoyamarev} such as magnetoelectric effect,\cite{Qi,Qi2,Nomura} spin torque and magnetization dynamics,\cite{Garate,Garate2,Yokoyama3,Yokoyama5,Mahfouzi,Nogueira,Semenov} magnetic domain walls,\cite{Tserkovnyak,Wickles,Ferreiros}
charge pumping,\cite{Ueda,Mahfouzi2}
magnetic heat transport,\cite{Yokoyama4}
spin rotation,\cite{Yokoyama1,Zhao}
spin and charge dynamics,\cite{Burkov,Schwab,Liu}
magnetotransport,\cite{Mondal,Yokoyama2,Wu,Zhang,Kong,Yang,Wang,Xia} and 
Majorana fermions (in superconducting heterostructures).\cite{Fu,Fu2,Akhmerov,Tanaka,Linder,Yokoyama6,Beenakker,Alicea} 
These phenomena are affected or effected by the exchange field of the F: The in-plane exchange field acts like a vector potential while the out-of-plane exchange field makes the Dirac fermions massive. On the other hand, the coupled spin and charge transport in metallic F's interfaces with TI's remains largely unexplored.\cite{Fischer} We study this transport here, in the weak-tunneling regime through the F/TI junction.

Motivated by theoretical predictions, there has also been intense interest on the magnetic proximity effect\cite{Henk,Henk2,Oroszlany,Luo,Eremeev,Menshov,Shen} and observation of  Dirac fermions coupled with exchange field.  Recently, signatures of Dirac fermions coupled to magnetization have been observed by doping Fe, Mn, Cr, Gd, or ferrocene into TI's,\cite{Chen,Hor,Okada,West,Song,Valla,Liu2,Cha,Checkelsky,Xu,Zhang2,Xu2,Yang2} depositing Fe or Co on the surface of TI's,\cite{Wray,Scholz,Honolka,Ye}  making junction of a TI with Fe, GdN or EuS,\cite{Vobornik,Kandala,Wei,Yang3} or intergrowth with Fe$_7$Se$_8$.\cite{Ji}

In this paper, we construct the diffusive theory for the coupled transport of the spin-dependent electron densities in the F and the charge density in the TI.
We derive the coupled spin and charge diffusion equations in layered metallic F/TI junctions, under current bias along the TI surface, which are perturbed by the tunneling across the interface. 
We analytically solve the diffusion equations to calculate the current-induced spin accumulation in the ferromagnet and junction magnetoresistance due to the F/TI interlayer tunneling.
It is found that due to the helical spin texture of the surface Dirac fermion, the spin accumulation  and the junction magnetoresistance depend on the angle between the magnetization and the current induced spin polarization on the surface of the TI.

\section{Model}

\begin{figure}[tbp]
\begin{center}
\scalebox{0.8}{
\includegraphics[width=10.0cm,clip]{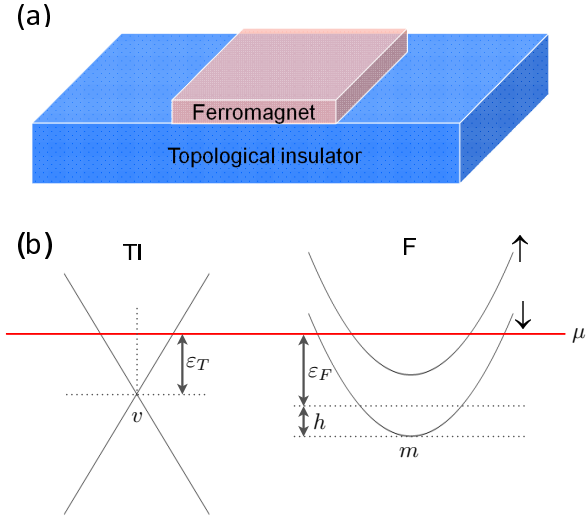}
}
\end{center}
\caption{ (Color online) (a) Schematic picture of the model. (b) Dispersions of the TI and F. The horizontal line denotes the Fermi level. }
\label{fig1}
\end{figure}

We consider an F/TI junction sketched in Fig.~\ref{fig1}(a). 
The Hamiltonian of the system reads $H = H_F  + H_T  + H'$, where
\begin{equation}
H_F=\sum\limits_\mathbf{k} {\hat{a}_\mathbf{k}^\dag  }\left(\frac{\hbar^2k^2}{2m}+\mathbf{h}\cdot\hat{\boldsymbol{\sigma}}-\varepsilon_F\right)\hat{a}_\mathbf{k}
\end{equation}
is the Hamiltonian of the ferromagnet. Here, $\mathbf{k}$ denotes 3D wave vector, $\mathbf{h}=h(\sin\theta\cos\varphi,\sin\theta\sin\varphi,\cos\theta)$ is the exchange field, $\hat{\boldsymbol{\sigma}}=(\hat{\sigma}_x,\hat{\sigma}_y,\hat{\sigma}_z)$ is a vector of Pauli matrices, and $\hat{a}^\dagger_\mathbf{k} = (a^\dagger_{\mathbf{k}\uparrow}, a^\dagger_{\mathbf{k}\downarrow})$ is the spin-1/2 creation operator.  We suppose that the magnetization is uniform. 
The Hamiltonian on the unperturbed surface of the topological insulator reads 
\begin{equation}
H_{T}  = \sum\limits_\mathbf{k} {\hat{b}_\mathbf{k}^\dag  } \left[ {\hbar v\left( {k_y\hat{\sigma}_x  - k_x\hat{\sigma}_y } \right) - \varepsilon _T } \right]\hat{b}_\mathbf{k}\,,
\end{equation}
where the 2D wave vector $\mathbf{k}$ is parallel to the interface.
The dispersions of the TI and F are shown in Fig.~\ref{fig1}(b), where the horizontal line denotes their common equilibrium Fermi level.

The F and TI are coupled by the tunneling Hamiltonian
\begin{equation}
H'= \sqrt{\frac{\hbar \gamma}{\pi}}\sum\limits_{\mathbf{k},\mathbf{k}'} {\hat{a}_\mathbf{k}^\dag  } \hat{b}_{\mathbf{k}'}+ {\rm H.c.}\,,
\end{equation}
where $\sqrt{\frac{\hbar \gamma}{\pi}}$ denotes the tunneling amplitude. It is assumed that the interface between F and TI is rough, such that, upon transmission through the interface, the momentum of the electron is scrambled while the spin and the energy are conserved. 
This spin-conserving form of tunneling would generally need to be extended to account for details of atomistic matrix elements for the states composing the Dirac electron band, in order to construct a more quantitative theory. We believe, however, that our simple treatment gives rise to the appropriate symmetry-based phenomenological structure of the final diffusion equations.
We will be assuming the ambient temperature $T$ to be low, $k_BT\ll\varepsilon_F,\varepsilon_T>0$, and will focus on  small deviations from an equilibrium state.

Spin-resolved electron-number density (measured relative to the equilibrium state) in the F, $n_\sigma$ with
$\sigma =  \uparrow$ and $\downarrow$ for minority and majority spins, respectively, obeys the diffusion equation:\cite{Johnson,Son,Valet,Hershfield}
\begin{equation}
\frac{\partial n_\sigma }{\partial t} = D_\sigma \nabla ^2 n_\sigma  - \left( \frac{n_\sigma}{\tau_\sigma} - \frac{n_{\bar{\sigma}}}{\tau_{\bar{\sigma}}} \right) -i'_\sigma \,,
\label{FD}
\end{equation}
where $i'_\sigma$ is the spin-$\sigma$ particle flux from the F into the TI across the F/TI interface and $\bar{\sigma}=-\sigma$.
According to equilibrium considerations, the spin-flip rates are related by $\rho_\uparrow/\tau_\uparrow=\rho_\downarrow/\tau_\downarrow\equiv\Gamma$, where $\rho _\sigma$ is the Fermi-level spin-$\sigma$ density of states in the F.
Electron density on the surface of the TI, $n$, in general obeys the continuity equation:
\begin{equation}
\frac{{\partial n }}{{\partial t}} = - \nabla \cdot {\bf{j}}+(i'_\uparrow+i'_\downarrow)\,.
\label{TD}
\end{equation}
Due to the strong spin-orbit coupling in the TI, we do not separate diffusion into two spin components. In lieu of the diffusive relations for $n_\uparrow$ and $n_\downarrow$ in the F, we have those for $n$ and $\mathbf{j}=(j_x,j_y)$ in the TI. The nonequilibrium spin density, as a hydrodynamic quantity, is thus replaced in the TI by the current density (which is physically motivated by the helicity of electron transport).

To the lowest order in tunneling, the spin-dependent interlayer currents can be calculated using the Fermi's golden rule: 
\begin{equation}
i'_\sigma  = \frac{{2\pi }}{\hbar }\sum\limits_{\mathbf{k},\mathbf{k}', \sigma} {| {\langle \hat{b}_{\mathbf{k}'} H' a_{\mathbf{k}\sigma}^\dagger\rangle_0 }|^2 (f_{F,\mathbf{k}\sigma}-f_{T,\mathbf{k'}})\delta (\varepsilon _{\mathbf{k}\sigma }    - \varepsilon _{\mathbf{k}'} )}\,.
\label{fgr}
\end{equation}
Here, $f_T$ and $f_F$ are the associated distribution functions; $\varepsilon_\mathbf{k}  = \hbar vk - \varepsilon _T$ and $\varepsilon _{\mathbf{k}\sigma }  = (\hbar k)^2/2m + \sigma h - \varepsilon _F$ are the corresponding energy eigenvalues; and the expectation value $\langle\dots\rangle_0$ is taken with respect to the absolute vacuum. In the state of global thermal equilibrium, the electron distributions are given by the Fermi-Dirac distribution function: $f_0(\varepsilon)=[e^{\varepsilon/k_BT}+1]^{-1}$. The nonequilibrium particle densities in the F and TI can be accounted for by simply shifting their chemical potentials by $\mu= n/\rho$ and $\mu_\sigma= n_\sigma/\rho_\sigma$, respectively, where $\rho$ is the Fermi-level density of states in the TI.

In the dilute limit of scattering impurities, the particle-current density $\mathbf{j}$ in the $xy$ plane on the surface of the TI results in a shift $\delta\mathbf{k}$ of the electron distribution function in the momentum space:
\begin{equation}
\mathbf{j}= \frac{{vk}}{{4\pi }} \delta\mathbf{k}\,,\,\,\,f_T=f_0-\frac{\partial f_0}{\partial \mathbf{k}}\cdot\delta\mathbf{k}\,.
\end{equation}
Here, $k$ is the Fermi wave number in the TI. 

\begin{widetext}
\section{Results}
\subsection{Diffusion equations}

The tunneling matrix elements entering in Eq.~\eqref{fgr} are determined by the spinor eigenfunctions 
\begin{equation}
\left| \mathbf{k} \right\rangle  = \frac{1}{{\sqrt 2 }}\left( {\begin{array}{*{20}c}
   i  \\
   { e^{i\phi } }  \\
\end{array}} \right)\,,\,\,\,
\left| \uparrow \right\rangle  = \left( {\begin{array}{*{20}c}
   {\cos \frac{\theta}{2}}  \\
   {e^{i\varphi } \sin \frac{\theta}{2}}  \\
\end{array}} \right)\,\,\,{\rm and}\,\,\,
\left| \downarrow \right\rangle  = \left( {\begin{array}{*{20}c}
   { - \sin \frac{\theta}{2}}  \\
   {e^{i\varphi } \cos \frac{\theta}{2}}  \\
\end{array}} \right)\,,
\end{equation}
in the TI and F, respectively, where $\phi  = \tan^{-1}(k_y /k_x)$ is the polar angle of $\mathbf{k}$ in the TI. We find (absorbing the ferromagnet's volume into $\gamma$)
\begin{equation}
i'_\sigma = \gamma\int \frac{k^2 dk}{2\pi ^2}\frac{k'dk'd\phi }{(2\pi )^2}\left[1 + \sigma \sin \theta \sin (\phi  - \varphi )\right]\left( {f_F  - f_T } \right)\delta(\varepsilon _{k\sigma }  - \varepsilon _{k'} )
  = \gamma\left[ { \frac{{\sigma\rho _\sigma \sin \theta }}{{v}} {\bf{e}}_\varphi   \cdot {\bf{j}} + \rho n_\sigma - \rho_\sigma n} \right]\,,
\label{js}
\end{equation}
where ${\bf{e}}_\varphi   = (\sin \varphi ,-\cos \varphi )$.

Substituting spin-dependent tunneling current density \eqref{js} into Eq.~\eqref{FD} complements the diffusion equation in the F with the tunneling leakage current, which is itself expressed in terms of the consistent hydrodynamics quantities. In order to complete our diffusion theory, we, however, still need to relate the current on the surface of the TI, $\mathbf{j}$, to the out-of-equilibrium densities, $n$ and $n_\sigma$, which we proceed to do in the following.

Driving the TI electrons by a spatially homogeneous in-plane electric field ${\bf{E}}$, in the presence of the interlayer tunneling, the current ${\bf{j}}$ on the surface of the TI obeys the following scattering relation:
\begin{eqnarray}
\frac{{\partial {\bf{j}}}}{{\partial t}} =  - \frac{\mathbf{j}-g\mathbf{E}/e}{\tau } + \left. {\frac{{\partial {\bf{j}}}}{{\partial t}}} \right|_t\,,
\end{eqnarray}
where $\tau$ is the transport mean free time, $e<0$ is the electron charge, and $g$ is the electrical conductivity, which is given, according to the Einstein relation, by $g = e^2 D\rho$, in terms of the diffusion coefficient $D=v^2\tau/2$. The last term on the right hand side represents the tunneling contribution to the time derivative of the current. 
At frequencies $\omega\ll1/\tau$, this reduces to 
\begin{equation}
 {\bf{j}} =  eD\rho {\bf{E}} + \tau \left. {\frac{{\partial {\bf{j}}}}{{\partial t}}} \right|_t
  \equiv {\bf{j}}_d  + {\bf{j}}_t \,,
\label{j}
\end{equation}
where the second term stands for the tunneling contribution to the planar TI current. In general, $-e {\bf{E}}$ should be replaced with the electrochemical potential gradient, $-e {\bf{E}}\to\boldsymbol{\nabla}\mu$, which also captures the diffusive contribution to the current: $\mathbf{j}_d=-D\boldsymbol{\nabla} n$. The tunneling contribution to the current is also evaluated using the Fermi's golden rule, giving
\begin{eqnarray}
\left.{\frac{{\partial {\bf{j}}}}{{\partial t}}} \right|_t  = \frac{{2\pi }}{\hbar }\sum\limits_{\mathbf{k},\mathbf{k}',\sigma } {{\langle \hat{b}_{\mathbf{k}'} {\bm{v}} \hat{b}_{\mathbf{k}'}^\dagger\rangle_0 } | {\langle b_{\mathbf{k'}}H' a_{\mathbf{k}\sigma}^\dagger\rangle_0 }|^2 (f_{F,\mathbf{k}\sigma}-f_{T,\mathbf{k'}})\delta (\varepsilon _{\mathbf{k}\sigma }    - \varepsilon _{\mathbf{k}'} )}
=  - \gamma\left[ {\rho _ +  {\bf{j}} + \frac{{v\sin\theta}}{2} ( { \rho n_-  -\rho _ -  n} ){\bf{e}}_\varphi } \right]\,.
\end{eqnarray}
where $ v_i  = \frac{{\partial H_{T} }}{{\hbar \partial k_i }} (i=x, y)$, $\rho_\pm\equiv\rho_\uparrow\pm\rho_\downarrow$ and $n_\pm\equiv n_\uparrow\pm n_\downarrow$. (We will later also benefit from the definition $D_\pm\equiv D_\uparrow\pm D_\downarrow$.) The first term gives a correction of order $\mathcal{O}(\gamma \tau)$ to the diffusion coefficient $D$, which can be absorbed by a redefinition of $D$. Note that in the absence of spin-orbit interactions in the F, $\partial \mathbf{j}_\sigma/\partial t|_t  = 0$ there, which allowed us to use simple diffusive relations in Eq.~\eqref{FD}.

Substituting the current Eq. \eqref{j} into Eq.~\eqref{TD}, we obtain the diffusion equation in the TI (to the lowest order in tunneling):
\begin{equation}
\frac{{\partial n }}{{\partial t}} = -\boldsymbol{\nabla}\cdot\mathbf{j}_d - \nabla  \cdot {\bf{j}}_t  + \gamma\left[ \frac{{\rho _ -  }\sin \theta}{{v}} {\bf{e}}_\varphi   \cdot {\bf{j}}_d  + \rho n_+ -{\rho _ +  }n\right] ,
\end{equation}
which becomes, upon substituting the above explicit expressions for $\mathbf{j}_d$ and $\mathbf{j}_t$:
\begin{equation}
\frac{{\partial n }}{{\partial t}} = D\nabla ^2 n + \gamma\left[ \frac{\rho D\sin \theta}{{v}} {\bf{e}}_\varphi   \cdot \boldsymbol{\nabla} n_- - \frac{{2\rho _ -  D\sin \theta }}{{v}}{\bf{e}}_\varphi   \cdot \boldsymbol{\nabla} n +\rho n_+-\rho_+ n \right]\,.
\end{equation}
In the most general case, our final diffusion equations for the TI and F are written in terms of the full electrochemical potentials (including electrostatic contributions), $\mu$ and $\mu_\sigma$:
\begin{equation}
\frac{{\partial n }}{{\partial t}} = D\rho\nabla ^2 \mu  + \gamma\rho\left[ {\frac{D\sin \theta}{{v}} {\bf{e}}_\varphi   \cdot \left(\rho_\uparrow\boldsymbol{\nabla} \mu _ \uparrow-\rho_\downarrow\boldsymbol{\nabla} \mu _ \downarrow\right) - \frac{{2\rho _ -  D\sin \theta }}{{v}}{\bf{e}}_\varphi   \cdot \boldsymbol{\nabla} \mu  + \rho _ \uparrow  \mu _ \uparrow   + \rho _ \downarrow  \mu _ \downarrow  -\rho _ +  \mu} \right]
 \label{Dm}
\end{equation}
and
\begin{equation} 
\frac{{\partial n_\sigma  }}{{\partial t}} = D_\sigma  \rho _\sigma  \nabla ^2 \mu _\sigma   - \Gamma( \mu _\sigma - \mu _ {\bar{\sigma}} ) + \gamma\rho \rho _\sigma\left[ {  \frac{{\sigma D\sin\theta  }}{{v}}{\bf{e}}_\varphi   \cdot \boldsymbol{\nabla} \mu  +   {\mu  - \mu _\sigma  } } \right] \,.
\label{Dms}
\end{equation}
These equations constitute the central result of this paper.


\subsection{Static 1D case}
As an example of F/TI junctions, consider the static 1D case. The diffusion equations in the F then become
\begin{equation}
0 = D_\sigma  \rho _\sigma \partial _x^2 \mu _\sigma   -  \sigma\Gamma( \mu _ \uparrow  - \mu _ \downarrow ) +\gamma \rho \rho _\sigma \left[ {\frac{\sigma D \sin \theta \sin \varphi}{{v}} \partial _x \mu  + \mu  - \mu _\sigma  } \right]\,,
\end{equation}
and
that  on the surface of the TI is
\begin{equation}
0 = D\partial _x^2 \mu  + \gamma\left[ {  \frac{D\sin \theta \sin \varphi}{{v }}  (\rho _ \uparrow \partial _x \mu _ \uparrow   - \rho _ \downarrow \partial _x \mu _ \downarrow  ) - \frac{{2\rho _ -  D\sin \theta \sin \varphi}}{{v }} \partial _x \mu  +  \rho _ \uparrow  \mu _ \uparrow   + \rho _ \downarrow  \mu _ \downarrow -\rho _ +  \mu  } \right] \,.
\label{Dm1}
\end{equation}
The diffusion equations in the F can be combined to obtain the equation for the spin accumulation $\mu_s\equiv\mu_\uparrow-\mu_\downarrow$:
\begin{equation}
0 = \partial _x^2 \mu_s - \frac{\mu_s}{{\lambda ^2 }}+ \gamma\rho\left[ {\left( {\frac{1}{{D_ \uparrow  }} + \frac{1}{{D_ \downarrow  }}} \right)\frac{D\sin \theta \sin \varphi}{{v }} \partial _x \mu  + \left( {\frac{1}{{D_ \uparrow  }} - \frac{1}{{D_ \downarrow  }}} \right)\mu  - \frac{{\mu _ \uparrow  }}{{D_ \uparrow  }} + \frac{{\mu _ \downarrow  }}{{D_ \downarrow  }}} \right]\,,
\label{Dms1}
\end{equation}
where $\lambda$ is the spin-diffusion length given by 
\begin{equation}
\frac{1}{{\lambda ^2 }} =\Gamma\left(\frac{1}{\rho_\uparrow D_\uparrow}+\frac{1}{\rho_\downarrow D_\downarrow}\right)\,.
\end{equation}
For the internal consistency of our quasi-1D treatment of the diffusion in the F, we need the F layer to be thinner than $\lambda$. The charge transport in the F is governed by the following relation:
\begin{equation}
0 =  (D_ \uparrow  \rho _ \uparrow \partial _x^2 \mu _ \uparrow   + D_ \downarrow  \rho _ \downarrow \partial _x^2 \mu _ \downarrow  ) + \gamma\rho\left[ {\frac{{D\rho _ - \sin \theta \sin \varphi}}{{v }} \partial _x \mu  +\rho _ +  \mu  - \rho _ \uparrow  \mu _ \uparrow   - \rho _ \downarrow  \mu _ \downarrow  } \right]\,.
\label{Dmc1}
\end{equation}
\end{widetext}

We will in the following solve the static 1D equations, Eqs.~\eqref{Dm1}, \eqref{Dms1}, and \eqref{Dmc1}, iteratively, to first order in $\gamma$. In the absence of tunneling, i.e., $\gamma=0$, the general solutions are 
\begin{align}
\mu_0  &= Ax + B\,,\nonumber\\
\mu_s ^0   &= ae^{x/\lambda }  + be^{ - x/\lambda }\,,\nonumber\\
\mu_c ^0&=\mu_0+cx + d-\frac{P\mu^0_s}{2}\,,
\end{align}
where $P\equiv(D_ \uparrow  \rho _ \uparrow-D_ \downarrow  \rho _ \downarrow)/(D_ \uparrow  \rho _ \uparrow+D_ \downarrow  \rho _ \downarrow)$ is the F conductivity polarization (according to the Einstein's relation) and $\mu_c\equiv(\mu_\uparrow+\mu_\downarrow)/2$ is the spin-averaged chemical potential.
Here, $a, b, c, d, A$, and $B$ are yet unknown constants, which should be determined by the boundary conditions. 
The solutions to the first order in $\gamma$ are then obtained by inserting these zeroth-order solutions into the terms proportional to $\gamma$ in the diffusion equations, which produces $\mathcal{O}(\gamma)$ source terms. After some algebra, we thus find solutions to the first order in $\gamma$:
\begin{widetext}
\begin{align}
\mu _ s =& ae^{x/\lambda }  + be^{ - x/\lambda } + \frac{\gamma\rho\lambda ^2}{D_\uparrow D_\downarrow} \nonumber\\
&\times\left[ {fe^{x/\lambda }  + ge^{ - x/\lambda }  +  \frac{{DD_+\sin \theta \sin \varphi}}{{v}}A +D_-(cx + d) + \frac{D_++PD_-}{4}\lambda x\left( ae^{x/\lambda }  - be^{ - x/\lambda }\right)} \right]\,,\nonumber\\
 \mu_c=& \mu_0+cx + d - \frac{\gamma \rho}{2(D_\uparrow\rho_\uparrow+D_\downarrow\rho_\downarrow)}\nonumber\\
 &\times\left[
 ux + v + \frac{{D\rho_- \sin \theta \sin \varphi}}{{v }}Ax^2   - \rho_+  \left( {\frac{c}{3}x^3  + dx^2 } \right)  
  + (P\rho_+-\rho_-){ \lambda ^2 \left(ae^{x/\lambda }  + be^{ - x/\lambda } \right) }
\right]-\frac{P\mu_s}{2}\,,\nonumber\\
 \mu  = &Ax + B + \frac{\gamma}{2D}\nonumber\\
 &\times\bigg[
  Cx + E- \frac{{D\sin \theta \sin \varphi }}{v}\left\{ {\left(\rho_+-P\rho_-\right) \lambda \left(ae^{x/\lambda }  - be^{ - x/\lambda } \right) + {\rho _ -  (c-A)}x^2 } \right\} \nonumber\\
  &\hspace{0.7cm} - \rho _ +  \left( {\frac{c}{3}x^3  + dx^2 } \right) 
  + (P\rho_+-\rho_-)  \lambda ^2 \left(ae^{x/\lambda }  + be^{ - x/\lambda } \right) 
  \bigg]\,.
\label{mmm}
\end{align}
Here, $f,g, u, v, C$, and $E$ are to be determined by the boundary conditions (at order $\gamma$).
\end{widetext}

In order to impose specific boundary conditions, suppose now that a ferromagnetic layer is attached to the TI along $-W/2<x<W/2$. Since the ferromagnet is terminated at $x=-W/2$ and $W/2$, we require vanishing of the diffusive spin-dependent planar fluxes normal to the boundaries, such that $\partial _x \mu_c = \partial _x \mu _ s = 0 $ at $x=\pm W/2$. 
Applying a uniform 2D particle flux $j$ to the TI surface in the $x$ direction, we have $j=-D\rho \partial _x \mu$, in the absence of tunneling, which determines $A=-j/D\rho$. We can, furthermore, set $B=0$ without loss of generality (absorbing it by a gauge potential shift, if necessary). Solving the first two of Eqs.~\eqref{mmm} then gives the spin accumulation in the F of the form:
\begin{align}
\mu_s  =& -4\gamma \lambda ^2j\left\{
\frac{\sin \theta \sin \varphi}{v (D_+-PD_-)}-\frac{D_ -}{D(D_+^2-D_-^2)}\right.\nonumber\\
&\hspace{2cm}\left.\times \left[x - \frac{\lambda\sinh(x/\lambda)}{\cosh(W/2\lambda) }\right]\right\}\,.
\end{align}
The spin accumulation, which is a nonequilibrium quantity, is proportional to the current $j$, as expected. The first term in the curly brackets is proportional to the $y$ projection of the F exchange field. This can be interpreted to result directly from the current-induced spin accumulation in the TI, which points along the $-y$ direction (when the current is flowing along the $x$ axis). Since the TI current is spin polarized helically even without the ferromagnet, this contribution is position-independent and persists even when $D_- \to 0$. The other contribution to $\mu_s$ is independent of the direction of the exchange field, and requires $D_-\neq0$, which means that it is due to nonequilibrium transport that is spin polarized by the ferromagnet. This contribution is position dependent, vanishing when $W\ll\lambda$ and growing linearly with $x$ when $W\gg\lambda$.

It should be remarked that $d$ is given by $d =  - j(\rho _-/v\rho\rho_+)\sin \theta \sin \varphi $, to the zeroth order in tunneling, reflecting the fact that the average chemical potential in the F can be shifted by spin injection. The tangential particle flux in the F is obtained as
\begin{equation}
\sum_\sigma j_\sigma=-\partial _x (D_ \uparrow  \rho _ \uparrow  \mu _ \uparrow   + D_ \downarrow  \rho _ \downarrow  \mu _ \downarrow  ) = \frac{\gamma \rho _ +}{2D}j\left( {\frac{{W^2 }}{4}} - x^2   \right) \,.
\end{equation}
The current in the F flows in the same direction as that on the surface of the TI and reaches its maximum value at the center of the junction. 

The chemical potential on the surface of the TI (setting $E=0$) can be expressed as
\begin{equation}
\frac{\mu}{\mu_0} = 1- \frac{\gamma}{2}\left[  \frac{\rho C}{j} - \frac{{\rho _ -  \sin \theta \sin \varphi}}{v}x  -\frac{{\rho _ + }}{{3D}}x^2  \right]\,,
\end{equation}
where $\mu_0=-jx/D\rho$. The continuity of the TI current \eqref{j} at junction boundaries, $x=\pm W/2$, at first order in $\gamma$, gives
\begin{equation}
\frac{\rho C}{j} = \frac{{\rho _ +  W^2}}{{4D}} - \frac{{4D\rho _ + }}{{v^2 }} - \frac{{2D\rho _ - ^2 \sin ^2 \theta \sin ^2 \varphi}}{{v^2 \rho _ + }}\,.
\end{equation}
Only when $\rho_- $ is nonzero, the chemical potential on the surface of the TI depends on the direction of the exchange field. 
Finally, the additional resistance $\delta R\equiv R-R_0$ of the TI/F junction (normalized to the bare TI resistance $R_0$) is found to be
\begin{align}
\frac{\delta R}{R_0} &= \frac{{\delta\mu ( - W/2) - \delta\mu (W/2)}}{jW/D\rho}\nonumber\\
& =\gamma\left[ { - \frac{{\rho _ +  W^2 }}{{12D}} + \frac{{2D\rho _ +  }}{{v^2 }} - \frac{{D\rho _ - ^2\sin ^2 \theta \sin ^2 \varphi }}{{v^2 \rho _ +  }} } \right]\,,
\end{align}
where $\delta\mu\equiv\mu-\mu_0$ is the chemical-potential shift due to tunneling. The magnetoresistance $\propto-h_y^2$ reaches maximum for $\mathbf{h}\perp\mathbf{y}$ and minimum for $\mathbf{h}\parallel\mathbf{y}$.
Note that this resistance respects the Onsager reciprocity for two-terminal systems. Namely, $\delta R\propto{\rm const}-h_y^2$ is invariant under time reversal: $\mathbf{h}\to-\mathbf{h}$. 
The normalized magnetoresistance is given by 
\begin{equation}
\frac{{R(\mathbf{h}\perp\mathbf{y}) - R(\mathbf{h}\parallel\mathbf{y})}}{R_0} = \frac{\gamma\rho _ - ^2 \tau}{2\rho_+} \,.
\end{equation}
For $\rho _ -  / \rho _ +   \sim 1$, $\hbar\gamma\rho_- \sim 10^{-3}$~eV, and $\tau  \sim 10^{ - 13}$ s, we estimate $\gamma\rho _ - ^2 \tau/(2\rho_+) \sim 0.1$.


\section{Conclusion}

In summary, we studied spin and charge diffusion in ferromagnet/topological-insulator junctions under current bias  on the surface of the topological insulator. The ferromagnet is assumed to be metallic.
The diffusive theory is constructed for the coupled transport of the spin-dependent electron densities in the ferromagnet and the charge density in the topological insulator. 
The diffusion equations which include the interlayer coupling and the current bias are derived based on the perturbative approach.
By solving the diffusion equations analytically, we obtained spin accumulation in the ferromagnet induced by the current and magnetoresistance due to the interaction between the magnetization and the current  flowing on the surface of the topological insulator. The dependences of the spin accumulation  and the junction magnetoresistance on the magnetization reflect the helical spin texture of the surface Dirac fermion.

\acknowledgments

This work is partly supported by a Grant-in-Aid for Young Scientists (B) (No. 23740236) and the ``Topological Quantum Phenomena" (No. 23103505) Grant-in-Aid for Scientific Research on Innovative Areas from theMinistry of Education, Culture, Sports, Science, and Technology (MEXT) of Japan, FAME (an SRC STARnet center sponsored by MARCO and DARPA), the NSF under Grant No. DMR-0840965, and Grant No. 228481 from the Simons Foundation.

%


\end{document}